\documentclass[reprint,
superscriptaddress,
nofootinbib,
 amsmath,amssymb,
 aps,
pra,
a4paper,twocolumn, floatfix
]{revtex4-2}
\usepackage[nobiblatex]{xurl} 
\usepackage{dsfont}
\usepackage{booktabs}
\usepackage{pgfplots}
\usepackage{subcaption}
\usepackage[labelfont=bf,
   justification=Justified,
   format=plain]{caption}
\usepackage{physics}
\usepackage{graphicx}
\usepackage{dcolumn}
\usepackage{comment}
\usepackage{bm}
\usepackage[hidelinks]{hyperref}
\usepackage{xcolor}
\usepackage[version=4,arrows=pgf-filled,
textfontname=sffamily,
mathfontname=mathsf]{mhchem}
\hypersetup{
    colorlinks,
    linkcolor={red!50!black},
    citecolor={blue!50!black},
    urlcolor={blue!80!black}
}
\usepackage[normalem]{ulem}

\newcommand{\shiftleft}[2]{\makebox[0pt][r]{\makebox[#1][l]{#2}}}

\date{\today}

\begin{document}

\title{Spectral Gap Estimation via Adiabatic Preparation}
\author{Davide Cugini}
 \email{davide.cugini01@universitadipavia.it}
 \author{Francesco Ghisoni}
 \email{francesco.ghisoni01@universitadipavia.it}
 \author{Angela Rosy Morgillo}
 \email{angelarosy.morgillo01@universitadipavia.it}
\author{Francesco Scala}
 \email{francesco.scala01@universitadipavia.it}
\affiliation{%
Dipartimento di Fisica, Universit\`a degli Studi di Pavia, via A. Bassi 6, 27100 Pavia (Italy)
}%

\begin{abstract}
    Estimating energy gaps, i.e. the energy difference between two different states, in quantum systems is crucial for understanding their properties. Conventionally, spectral gap estimation relies on independently computing the ground-state and first-excited-state energies and then taking their difference.
    This work introduces an alternative procedure for estimating spectral gaps on \emph{digital quantum devices} using the Adiabatic Preparation technique to create a specific superposition state. 
    The expectation values of observables measured on such a state exhibit time-dependent fluctuations which, through a fitting process, can be used to estimate the energy gap.
    We successfully test our method on the 1D and 2D Ising models, and \ce{H2} and \ce{He2} molecules, implementing relatively shallow circuits both on noiseless and noisy simulators. The robustness of the approach is corroborated by additional experiments on the real IonQ Aria device for the 1D Ising model up to 20 qubits, demonstrating the applicability of the proposed method for currently available digital quantum devices and paving the way for more complex energy gap calculation requiring deeper circuits in the fault-tolerant era to come.
\end{abstract}

\maketitle

\begin{figure*}
    \centering
    \includegraphics[trim={1.8cm 10cm 2.cm 9cm},clip,width=.85\textwidth]{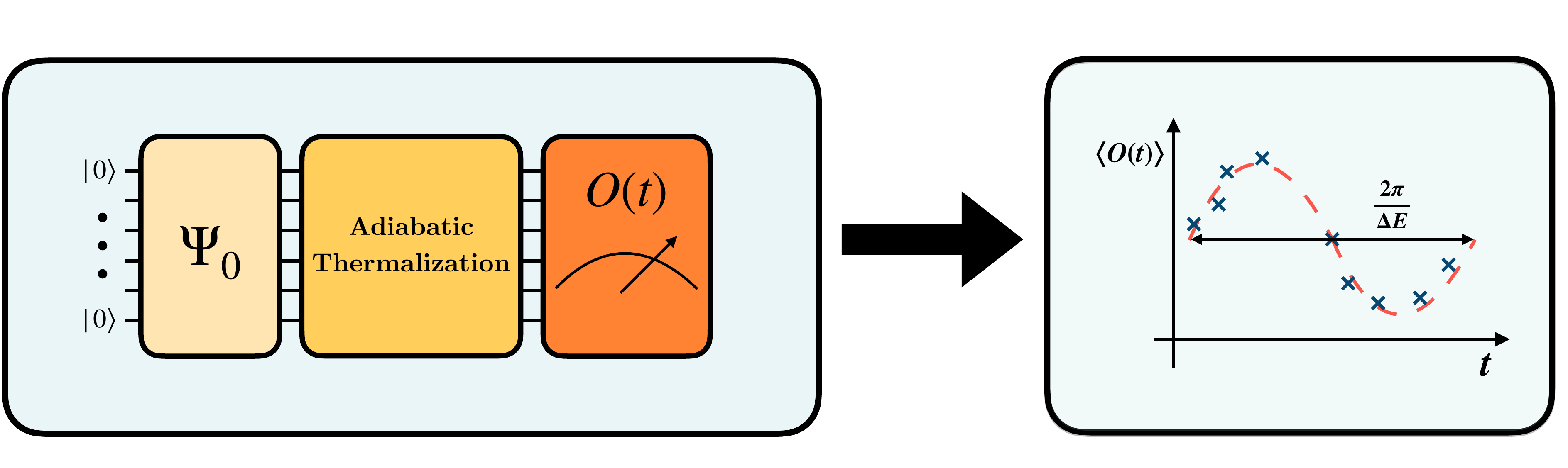}
    \caption{Schematic representation of the proposed methodology. Initially, a quantum circuit generates the superposition state $\ket{\Psi_0}$ of two eigenstates of the auxiliary Hamiltonian $H_0$. Through Adiabatic Preparation this state is evolved to the $\ket{\Psi}$ state of the system Hamiltonian $H$. Sequential measurements of the chosen observable $O(t)$ at different time steps are recorded. Finally a fitting process is used to determine the spectral gap. The obtained value is compared to the benchmark spectral gap.
 }
    \label{fig:scheme}
\end{figure*}

\section{Introduction}

The estimation of the spectral gap, 
defined as the energy difference between the ground eigenstate 
and the first excited state of a quantum system, 
is fundamental in both condensed matter physics~\cite{somma2013spectral, osterkorn2023gap, cubitt2015undecidability, gosset2016correlation, can2019spectral, arad2017rigorous, hastings2007area, landau2015polynomial,sachdev2011quantum,hastings2006spectral,kohn1964theory,szymanski2022universal,CHUNG2025110486,rai2025spectral},  and quantum chemistry~\cite{tiwary2016spectral,bredas2014mind,griffith1957ligand,lupo2023two, deshpande2022importance}. 
In the first case, the spectral gap determines key aspects of a material’s behavior, particularly at low temperatures, and provides insights into its phase transitions~\cite{sachdev2011quantum} and stability~\cite{hastings2006spectral}. From the point of view of conductivity~\cite{kohn1964theory}, systems with a large spectral gap tend to be insulators,  while conductors and metals are characterized by the absence of a gap at the Fermi level. Moreover, materials with small or vanishing gaps near certain critical points can exhibit dramatic changes in their physical properties, signifying phase transitions and critical phenomena~\cite{szymanski2022universal, CHUNG2025110486}.  
Among the many energy gaps that can be computed~\cite{bredas2014mind}, the spectral gap between the highest occupied molecular orbital and the lowest unoccupied molecular orbital is a crucial quantity in quantum chemistry for determining a molecule’s stability and reactivity~\cite{griffith1957ligand}
and affects various chemical processes, including catalysis, bonding, and photochemistry, where energy absorption and transfer depend on these excitation energies~\cite{lupo2023two}.

Despite its fundamental importance, calculating the spectral gap, especially for complex or large quantum systems, poses significant computational challenges on a classical computer. Recent results demonstrate that the problem matches the complexity of the most challenging tasks in quantum verification, confirming its QMA-hardness even under restricted settings~\cite{Yirka2025}. 
For fermionic systems, the occurrence of the so-known \emph{sign problem} can hinder the computation, particularly for systems with strong interactions or at low temperatures~\cite{troyer2005computational}. 
Moreover, 
the exponential growth of the space of configuration dimension
with the size of the system implies that exact solutions quickly become unfeasible, requiring the use of approximated methods~\cite{cao2022progress}.

With the first ideas dating back to the 1980’s, the field of quantum simulation has sought ways to learn how to control quantum devices, analog or digital, to simulate the dynamics of quantum mechanical systems. These devices have allowed researchers to develop novel quantum algorithms that can potentially overcome the fundamental issues encountered by classical computers. 
However,
currently available digital devices are both noisy and limited in the number of qubits, defining the so-called Noisy Intermediate-Scale Quantum (NISQ) era~\cite{preskill2018nisq}. While these devices are still far from ideal for large-scale, fault-tolerant quantum computation, they offer a unique opportunity for tackling certain problems in quantum physics that are otherwise intractable on classical computers. Within this framework, researchers have explored various techniques to leverage NISQ devices for estimating energy gaps. Methods such as imaginary time propagation~\cite{leamer2023spectral}, robust phase estimation~\cite{russo2021robustqpe}, and Bayesian phase difference estimation~\cite{GNATENKO2022127843} have been proposed. 

In this work, building on the previous proposal of Ref.~\cite{GNATENKO2022127843}, 
we present an improved method
exploiting the expectation-value fluctuations of physical observables
to directly estimate energy gaps, 
eliminating the need to compute individual energy eigenvalues.
Our starting point is
the superposition state 
\begin{equation}
\label{eq:SGS}
\ket{\Psi(i,j)} \equiv \frac{1}{\sqrt{2}}\left(\ket{\Omega_i}  + \ket{\Omega_j} \right),
\end{equation}
where $\ket{\Omega_i}$ and $\ket{\Omega_j}$
respectively are the $i$-th and $j$-th eigenstates
of the system Hamiltonian $H$.
Here, we exploit the Adiabatic Preparation (AP)~\cite{born1928beweis} to initialize $\ket{\Psi}$ states on a digital quantum computer.
Although the two eigenstates are not known analytically,
they can be prepared on a quantum computer through AP
if the adiabatic theorem conditions hold~\cite{jansen2007bounds}, which requires the time evolution operator for the Hamiltonian to have no energy level crossings. We note that, to our knowledge, there is no other approach based on the Variational Quantum Eigensolver (VQE) to prepare the state of Equation~\eqref{eq:SGS}.
In this regime, it is possible to utilize
the time fluctuations of observables 
evaluated on the $\ket{\Psi}$ state
to obtain an estimate of the energy gap 
\begin{equation}
    \Delta E_{j,i} = E_i-E_j \ ,
\end{equation}
where $E_i$ and $E_j$ are the energy eigenvalues associated to $\ket{\Omega_i}$ and $\ket{\Omega_j}$,
without the need to compute each energy individually.
We highlight that this procedure can be directly applied on a digital quantum computer to estimate the energy gaps between all possible energy levels, provided the quantum adiabatic theorem holds. This stands in contrast to quantum annealers, which require gap-specific engineering~\cite{matsuzaki2021direct}. We validate the procedure by applying it on both quantum simulators and real quantum hardware for different problems. Even though the simulated systems are limited in size, we point out that the experimental results are intended as a proof of principle showing the applicability of the method on currently available hardware.

This article is organized as follows. In Section~\ref{subsec:background} we introduce the technical background required for this work.
In Section~\ref{subsec:SGestimation} we present our procedure, 
showing its derivation and underlying principles.
In Section~\ref{sec:results} 
we report the numerical results 
of our method applied for the spectral gap estimation
of the Ising model,
on a classical computer simulator,
followed by the outcomes 
obtained with the real hardware IonQ Aria. 
It follows a detailed description of the spectral gap estimation
for the \ce{H2} and \ce{He2} molecules
run on a classical computer simulator.
Finally, in Section~\ref{sec:discussion}, 
we provide a comprehensive summary 
and draw conclusions from our findings.  

\section{Methods}
\label{sec:method}
In this section, we present the underlying theory needed for our study, including adiabatic state preparation and the Trotter-Suzuki decomposition, and describe how these techniques are applied to estimate energy gaps.
\subsection{Background}
\label{subsec:background}

In this section, we present the technical background that will be used throughout the paper. Specifically, we provide knowledge about Adiabatic Preparation (AP)~\cite{born1928beweis, jansen2007bounds} needed to initialize the register in the desired superposition states defined in Equation~\eqref{eq:SGS},
and about the \emph{Trotter-Suzuki} formula for time evolution~\cite{trotter1959product}, which will be employed to implement both AP and time evolution of an observable.

AP is a common technique in quantum simulations that aims at preparing the $n$-th eigenstate of a complex Hamiltonian $H$. To do so it first initializes a quantum register in the $n$-th eigenstates of an auxiliary Hamiltonian $H_0$, which is typically easy to prepare, e.g. $H_0 = \mathrm{diag}(H)$. Then such eigenstate is evolved through a different time-dependent Hamiltonian $\Tilde{H}$. By ensuring an (almost) adiabatic evolution, it is guaranteed that the final state will be the $n$-th eigenstate of $H$~\cite{jansen2007bounds}. 
The evolution to prepare the eigenstates of $H$ can be performed via a time-dependent Hamiltonian \begin{equation}\label{eq: AP Hamiltonian}
\Tilde{H}\left(\frac{t}{\tau}\right) =H_0 + \frac{t}{\tau}\left(H-H_0\right) \, ,
\end{equation}
where $t$ is the time variable
and $\tau$ is the duration of the adiabatic process,
such that $\Tilde{H}(0) = H_0$ and $\Tilde{H}(1) = H$.
The time evolution operator associated with $\Tilde{H}$ is
\begin{equation}\label{eq: time evolution operator}
    U_\tau\left(\frac{t}{\tau}\right) = \mathcal{T}\mathrm{exp}\left[-i \tau\int_0^{t/\tau} ds \Tilde{H}(s) \right] \, ,
\end{equation}
where $\mathcal{T}$ is the time ordering operator.
In the limit of $\tau \to \infty$ 
the \textit{adiabatic theorem} guarantees that,
under general hypothesis~\cite{born1928beweis, jansen2007bounds},
the $n$-th eigenstate $\ket{\omega_n}$ of $H_0$
is evolved by adiabatic evolution into
\begin{equation}
    \lim_{\tau \to \infty}U_\tau(1)\ket{\omega_n} = \ket{\Omega_n},
\end{equation}
where $\ket{\Omega_n}$ is the $n$-th eigenstate of $H$.
We highlight that the main condition for the success of the AP protocol is determined by the spectrum of $H\left( \frac{t}{\tau}\right)$. In particular, if the $n$-th eigenstate of $H(\frac{t}{\tau})$ showcases a non-vanishing gap with the rest of the spectrum during the entire adiabatic process then the error of the prepared eigenstate will decay at least exponentially with $t$. Otherwise, $\ket{\omega_n}$ is not guaranteed to converge to $\ket{\Omega_n}$. From these considerations it is clear that level crossings are the fundamental limitation to the effectiveness of AP protocol.
In what follows, 
we make use of the Branched-Subspace Adiabatic Preparation (B-SAP) 
algorithm for the state preparation,
which relaxes the strict energy-gap conditions required in conventional adiabatic protocols~\cite{cugini2025bsap}.
In particular, 
it extends the applicability of the adiabatic theorem to situations in which the initial Hamiltonian $H_0$ has a degenerate spectrum.

The time evolution of $\Tilde{H}$ cannot be straightforwardly implemented on digital quantum computers. 
Instead, one can approximate such continuous dynamics with a sequence of discrete steps by exploiting the so called Trotter-Suzuki formula (also known as Trotterization)~\cite{trotter1959product}.
The latter comes in handy not only to perform the AP of the superposition states, but also enables the subsequent simulation of the time evolution of observables $O(t)$ in the Heisenberg picture.
In particular, 
for the AP process defined by $\Tilde{H}$ in Equation~\ref{eq: AP Hamiltonian}, we can rewrite its associated time-evolution (Equation~\eqref{eq: time evolution operator}) as
\begin{equation}
    U_\tau(1) =  \mathcal{T}\mathrm{exp}\left[ -i\frac{\tau}{Q} \sum_{q=1}^Q \Tilde{H}\left( \frac{q}{Q+1}\right)\right] \, ,
\end{equation}
where $\tau/Q$ defines the length of the discretization, often referred to as the time step, and $\Tilde{H}\left( \frac{q}{Q+1}\right)$
is the Hamiltonian at the $q$-th time step. 
Then, we can express the Hamiltonian as a sum of terms 
\begin{equation}\label{eq: H Pauli decomposition}
    \Tilde{H}(s) = \sum_ic_i(s)\Tilde{H}_i \, , \quad \quad c_i \in \mathbb{R} \, ,
\end{equation}
such that the unitary operators 
$U_i(\delta t) = e^{-i\delta t \Tilde{H}_i}$
can be implemented on the quantum register for a time step $\delta t$. In quantum computing a typical choice is to decompose $\Tilde{H}(s)$ into Pauli strings, which is convenient since they form a basis for the unitary operators on the $2^L$ dimensional Hilbert space of an $L$-qubit system, and it is well-known how to encode their exponential operators on a digital quantum computer~\cite{Tacchino2020AQT}.
Finally, we can apply the Trotter-Suzuki formula to obtain
\begin{align}
    U_\tau(1) 
    &= \mathcal{T} \,\prod_{q=1}^{Q}\prod_iU_i\left(-i\frac{\tau}{Q}\,c_i\left(\frac{q}{Q+1} \right)\right) \nonumber \\
    &\quad\quad+ \mathcal{O}\left(\frac{\tau^2}{Q}\right)  \,, 
\end{align}
that can be readily implemented on a quantum register.
A similar procedure can be applied to the time-independent target Hamiltonian $H$ to perform the time evolution of observables $O(t)$ in the Heisenberg picture.

For an efficient preparation of the $n$-th eigenstate, a system needs a sufficiently large gap between such state and the $n+1$-th eigenstate, as an excessively small gap can cause the AP to fail within a reasonable time frame. Hence, this can lead to prolonged evolution times and potential violations of the adiabatic theorem.

\subsection{Spectral gap estimation}
\label{subsec:SGestimation}
This section introduces the proposed methodology by building on the principles of the AP and the Trotter-Suzuki formula discussed in Section~\ref{subsec:background}. Our approach leverages the oscillatory behaviour in time of the expectation value of an observable $O(t)$, using its period to estimate the spectral gap. Validation of the methodology is provided through numerical results and experiments on real quantum hardware in Section~\ref{sec:results}.

Given a time-\emph{independent} operator $O$, we can define the time-\emph{dependent} operator $O(t)$ as
\begin{equation}
 O(t)   =  e^{iHt} O e^{-iHt},
\end{equation}
where $H$ is the Hamiltonian of a physical system.
The expectation value of the observable $O(t)$ calculated on the superposition state $\ket{\psi}$ (defined in Equation~\eqref{eq:SGS}) reads
\begin{equation}\label{eq: fluctuations}
\begin{split}
    \langle O(t) \rangle &= \frac{1}{2}\left( \bra{\Omega_i} O \ket{\Omega_i} + \bra{\Omega_j} O \ket{\Omega_j} \right) + \\
&\quad\quad + \mathcal{A}\,\mathrm{cos}\left(\Delta E_{ji}t + \theta\right),
\end{split}
\end{equation}
where $\mathcal{A} > 0$ and $\theta \in [0, 2\pi)$
are the polar coordinates
of the complex number $\bra{\Omega_j} O \ket{\Omega_i}.$ 
The procedure to find the energy gap consists of two main steps: the first involves calculating the value of the observable, Equation~\eqref{eq: fluctuations}, at $R$ different times $\{t_r\}_{r=1}^R$ to create the dataset $\mathcal{D} = \{\langle O(t_r) \rangle \}_{r=1}^R$
and the second step uses $\mathcal{D}$ to approximate the energy gap $\Delta E_{ji}$.

To obtain each $\langle O(t_r) \rangle$ one should: 
(i) prepare the desired superposition state $\ket{\Psi}$ on a quantum computer,
(ii) implement the time evolution $e^{-iHt_r}$ 
as a quantum circuit and 
(iii) compute the expectation value of a time-independent operator $O$ 
(see Figure~\ref{fig:scheme}). Then $\mathcal{D}$ can be used in two ways to approximate the energy gap: 
either by fitting it to a sinusoidal function and calculating the period, or by performing a Discrete Fourier Transform on it. 
Although the Discrete Fourier Transform is a valuable method, 
it only allows to consider a finite number $R$ of frequencies, 
equal to the size of $\mathcal{D}$,
requiring a high number of samples to obtain a precise estimation of the energy gap.
For this reason, in this work, we adopt the fitting approach as we can estimate $\langle O(t_r) \rangle$ only for a limited set of time-coordinate values due to limited access to real quantum devices.

Following the guidelines presented in Section~\ref{subsec:background}, 
the state preparation can be performed 
via the AP. 
This allows to evolve each $n$-th eigenstate $\ket{\omega_n}$ 
of an auxiliary Hamiltonian $H_0 $
into the corresponding $n$-th eigenstate of $H$ 
through the evolution operator $U_\tau$,
where $\tau > 0$ represents the adiabatic preparation time.
This guarantees that,
starting from the superposition
\begin{equation}
    \ket{\Psi_0} = \frac{1}{\sqrt{2}}\left(\ket{\omega_i}+ \ket{\omega_j} \right) \,,
\end{equation}
then 
\begin{equation}
    \lim_{\tau \to \infty}U_\tau\ket{\Psi_0} = \ket{\Psi}.
\end{equation}
Fixing a finite number of Trotter steps during the AP inevitably leads to imperfect state preparation. Despite the finite fidelity, the technique remains effective as long as the contributions from the other components of the state are negligible compared to the main terms. Theoretical insights into this phenomenon are reported in Appendix~\ref{appendix:imperfect sgs}.

It is important to note that,
as long as $\mathcal{A}>0$,
any observable $O$ can be used in Equation~\eqref{eq: fluctuations}
to estimate the energy gap.
\textit{A priori} knowledge of the system, such as numerical experiments or symmetry considerations, should be used to find the observable that maximizes $\mathcal{A}$, since this facilitates the fitting process. See Appendix~\ref{appendix:observable} for more details about the observable choice.

Observe that our procedure allows,
in principle, 
to estimate the energy gap between two arbitrary eigenstates of $H$.
Moreover, this is done without separately calculating the two energy eigenvalues.
Therefore, the presented technique is particularly suitable in those cases where the energy estimation is a demanding operation, e.g. in the case of Hamiltonians with many terms whose expectation value cannot be simultaneously estimated.
\section{Results}
\label{sec:results}
In this section, we present the numerical tests
with the developed procedure.  
Aiming at implementing the 
numerical studies on a digital quantum device,
we restrict the circuits to be composed of 
40 Trotter steps.
From now on we use the following notation
for the $2\times2$ identity operator 
and the Pauli matrices
\begin{equation}
    \sigma^0 = I \, , \quad \sigma^1 = \sigma^x \, ,
    \quad \sigma^2 = \sigma^y \, ,
    \quad \sigma^3 = \sigma^z \, .
\end{equation}

\subsection{Ising model}
\label{sec:Ising}
Our first test case is the $L$-sites Ising model \cite{huang2008statistical}
with Periodic Boundary Conditions (PBC), 
described by the Hamiltonian
\begin{equation}
    H = -\frac{J_1}{2} \sum_{\langle ij \rangle} \sigma^1_i \sigma^1_{j} - \frac{h_3}{2}\sum_{i}\sigma^3_i \, ,
    \label{eq:ising}
\end{equation}
where $\sigma_i$ is a Pauli matrix acting on the $i$-th site 
and the $\langle \cdot \cdot  \rangle$ symbol 
restricts the sum over the nearest neighbours.
In what follows, 
we restrict to the case 
of positive 
$J_1 $.
To find the spectral gap
we need to prepare the superposition state $\ket{\Psi(0,1)} = \frac{1}{\sqrt{2}} (\ket{\Omega_0}+ \ket{\Omega_1})$.

For this scope, we consider the auxiliary Hamiltonian
\begin{equation}\label{eq : initial Hamiltonina Ising}
    H_0 = -\frac{J_1}{2}\sum_{\langle ij \rangle } \sigma^1_i\sigma^1_{j} \, ,
\end{equation}
corresponding to Equation~\eqref{eq:ising} with $h_3=0$.
The ground-state subspace of $H_0$ is twofold degenerate; while any orthonormal basis could be chosen, we adopt $\ket{+}^{\otimes L}$ and $\ket{-}^{\otimes L}$, defined as:
\begin{equation}
    \ket{\pm} \equiv \frac{1}{\sqrt{2}}\left(\ket{0} \pm \ket{1} \right).
\end{equation}
Any linear combination of 
\(\ket{+}^{\otimes L}\) and \(\ket{-}^{\otimes L}\) is itself a ground state. 
Preparing the state \(\ket{\Psi_0}\), 
which is then adiabatically evolved into \(\ket{\Psi(0,1)}\), 
requires constructing the appropriate superposition 
in such a degenerate eigenspace.
This task can be carried out straightforwardly 
using the B-SAP algorithm~\cite{cugini2025bsap},
by introducing a parametrized unitary to be optimized variationally.
However, based on symmetry considerations (see Appendix~\ref{Ising SGS}), it is possible to bypass the variational procedure and analytically determine 
\begin{equation}
    \ket{\Psi_0(0,1)} =  \ket{+}^{\otimes L}.
\end{equation} 
All experiments conducted for the Ising model,
both in 1D and 2D, use the observable
\begin{equation}\label{eq:observableising}
    O =  \sigma^1 \otimes \left(\sigma^0\right)^{\otimes (L-1)}.
\end{equation}
Both analytical considerations and numerical experiments
seems to suggest that this observable maximizes $\mathcal{A}$ 
in Equation~\eqref{eq: fluctuations} 
for any Ising chain (1D case) and square lattice (2D case). 
The full investigation can be found 
in Appendix~\ref{appendix:observable}.

The presented experiments are a 
4-sites and 20-sites Ising chain with PBC.
It is important to note that
the circuit depth per Trotter step needed to simulate Ising models is independent of the number of sites $L$. 
In particular, all experiments 
involving Ising chains, 
including noiseless, noisy and real hardware, 
have a maximum depth of 80 2-qubit gates
once decomposed with IonQ 
native gates. 
More information can be seen in Appendix~\ref{appendix:ionq}.

The 40 Trotter steps
are divided in 15 adiabatic steps 
and 25 time evolution steps. We choose the 25 times $t$ for the evaluation of $O(t)$ 
to be Chebyshev-distributed 
to facilitate the fitting process~\cite{Rendon2024improvedaccuracy}. Specifically, to avoid missing the oscillations of $O(t)$, which could happen if uniformly distributed times coincided with the sinusoidal function's nodes, we choose times as Chebyshev nodes~\cite{chebyshev1853theorie}.
Each circuit
is simulated on a classical computer 
with 8192 shots.
We take the intrinsic error 
of the quantum state measurement 
into account as described in Appendix~\ref{sec: Intrinsic error}. 
All simulated results are compared
to a numerical benchmark which is 
calculated using direct diagonalization of 
the Hamiltonian $H$.

\begin{figure*}[t]
\centering
    \includegraphics[trim={.cm 0.cm 0cm 0cm},clip,width=.45\textwidth]{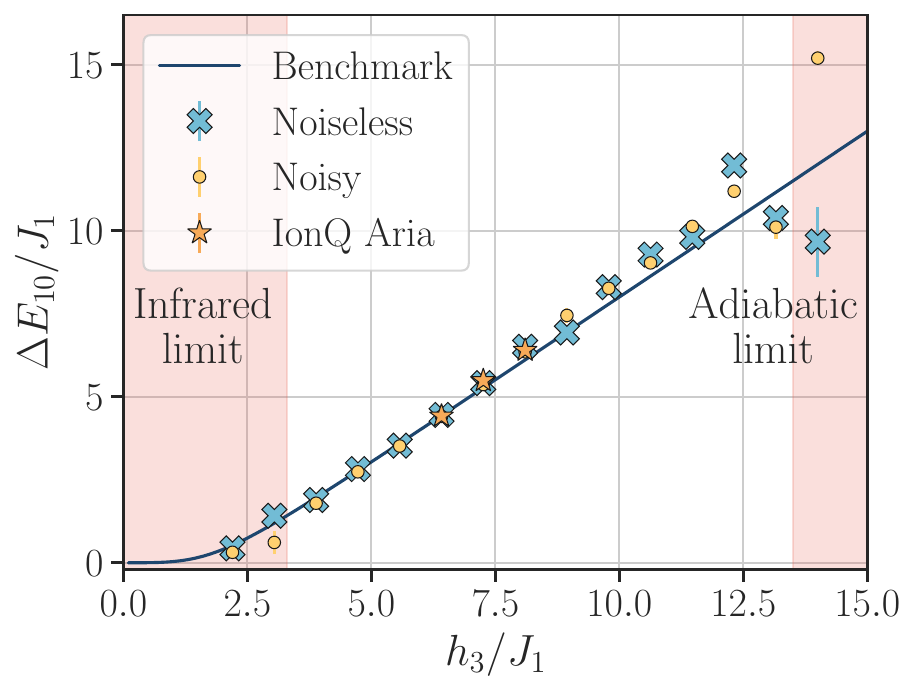}
    \shiftleft{8.4cm}{\raisebox{5.3cm}[0cm][0cm]{(a)}}
    \label{fig:spectralgaprunning4}
    \includegraphics[trim={.cm .cm 0.cm 0cm},clip,width=.45\textwidth]{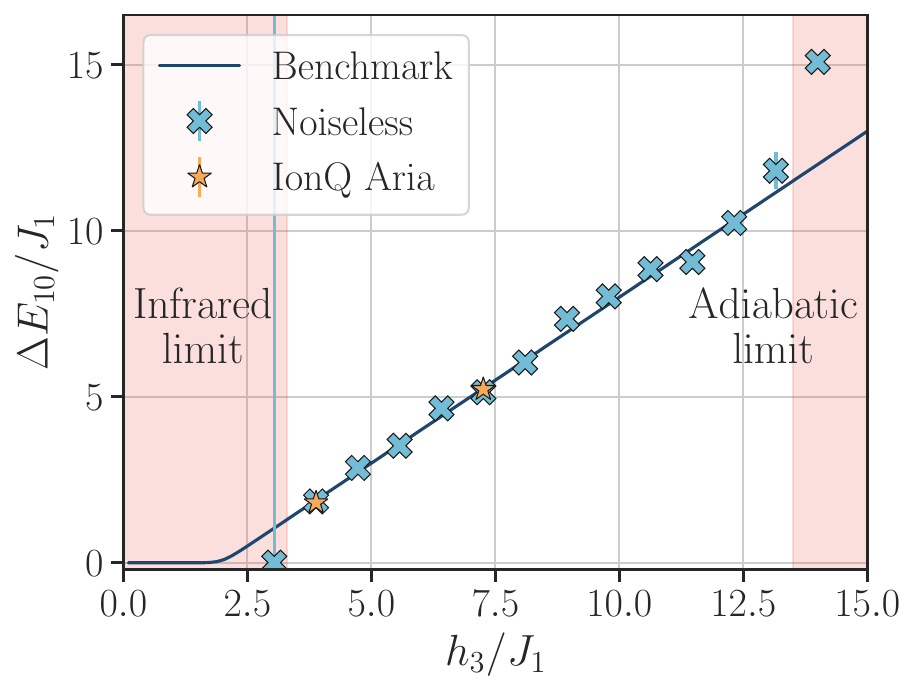}
    \shiftleft{8.4cm}{\raisebox{5.3cm}[0cm][0cm]{(b)}}
    \label{fig:spectralgaprunning20}
\caption{Spectral gap as a function of $h_3/J_1$ for a (a) $L=4$ and (b) $L=20$ Ising chain with PBC. The plots include simulations on a classical computer, showing both noiseless results and experiments on IonQ Aria 1.  The adiabatic and infrared limits are regions where the circuit depth used is insufficient. Classical simulations of the noisy hardware are also shown in panel (a). Error bars, shown but not always visible, represent a single standard deviation computed from the fit.\label{fig:Ising}}
\end{figure*}

The numerical results for the 4-site Ising chain, presented in Figure 
\ref{fig:Ising}a, 
show an excellent agreement
between the numerical benchmark 
and the noiseless estimated spectral gap for $L=4$. 
The graph illustrates the spectral gap 
dependence on the parameter $h_3/J_1$. 
The relative error between the noiseless 
numerical results and the benchmark is in the order of $10^{-2}$.
In Fig.~\ref{fig:Ising} we present numerical results for the Ising chain, showing the dependence of the spectral gap on the parameter $h_3/J_1$. For the 4-site case (Fig.~\ref{fig:Ising}a), the noiseless and noisy estimate closely matches the numerical benchmark, with a relative error of order $10^{-2}$ on average. This explains why error bars are not visible in general. The analysis extended to 20 sites (Fig.~\ref{fig:Ising}b) shows similarly strong agreement, with a relative deviation of order  $10^{-2}$, except for one point in the infrared limit where the estimated error is extremely large.

The infrared and adiabatic limits are regimes where our approach becomes unstable,
within the imposed number of Trotter steps. 
Even if one tries to balance the reduced number of Trotter steps with longer time steps, at a certain point, the Trotter-Suzuki approximation would no longer hold. 
In particular, in the infrared limit, the $O(t)$ frequency tends to zero 
and its estimation would require a huge amount of time steps for the time evolution. 
The adiabatic limit, instead, requires a long adiabatic time for a high fidelity state preparation. 
These limits were not estimated analytically: the red-shaded areas in Figures~\ref{fig:Ising} and \ref{fig:Ising 2D} highlight regions where experiments become inaccurate.
As a matter of fact, we noticed that the fidelity between the prepared superposition state and the spaces spanned by ground and first-excited states decreases, for increasing transverse field. In addition, in such experiments, we observe small periodic oscillations in the fidelity for which a theoretical explanation is missing at the time of the writing. These additional experiments are reported in Appendix~\ref{appendix:imperfect sgs}. 

Remarkably, our approach exhibits strong resilience 
when subjected to the noise model inspired by IonQ Aria 1. 
Specifically, the impact of thermal relaxation and readout are minimal due to the extended $T_1$ and $T_2$ parameters and small readout error, while the biggest error contribution is due to 2-qubit gate average fidelities (see Appendix~\ref{appendix:ionq} for more details on the hardware noise model). 
The average relative error between the numerical and the estimated gap has been determined to be of order $10^{-1}$.
Moreover, further investigation revealed that the procedure has also strong
resilience to a reduced number of shots (see Appendix~\ref{appendix:ionq}).

Analogous results are obtained for the 4-sites 2D Ising lattice 
and are reported in Appendix~\ref{subsec:Ising2d}. 

\subsubsection*{IonQ Aria Device}
The final test
for our procedure is carried out on the real IonQ Aria 1 device on 4- and 20-sites Ising chain. 
Due to resource constraints,
the conducted experiments 
used slightly different specifics compared
to the noiseless and noisy experiments. 
In addition, the number of experiments was severely limited by the availability of the device (almost always under maintenance) and bad fidelity calibrations since December 2024.

The energy gap for only three different 
$h_3/J_1$ values is found for 4 qubits and only two for 20 qubits experiments. 
These values are selected 
since they are far from both the infrared and adiabatic limits.
Secondly, only a subset of 10 times $t$ of the initial 25, 
are employed for the fitting of $O(t)$. 
Finally, 2500 shots (with error mitigation) 
are used instead of 8192 employed in the numerical 
simulations (without error mitigation).

The results, shown in Figure~\ref{fig:Ising}, reveal agreement between the noiseless simulation, the noisy simulation, and the IonQ Aria device. This witnesses the robustness of our approach to noise (both simulated and real). It is worth mentioning that the IonQ Aria device comes with integrated error mitigation~\cite{maksymov2023enhancing} for a minimum number of 2500 shots, which contributed to the consistent results obtained in our study. Moreover, thanks to noise mitigation results on real hardware are actually better than the ones obtained with noisy simulations as they have a larger $\mathcal{A}$ (see Appendix~\ref{appendix:ionq}).

\subsection{\ce{H2} and \ce{He2} molecules}
In this section, we present the results of our spectral gap estimation applied to the \ce{H2} and \ce{He2} molecules. 

The molecular electronic Hamiltonian $\mathcal{H}$ is expressed as 
\begin{equation}\label{eq: ce{H2} Hamiltonian}
    \mathcal{H} =  \sum_{pq} h_{pq} c_p^\dagger c_q  + \frac{1}{2}\sum_{pqrs} h_{pqrs} c_p^\dagger c_q^\dagger c_r c_s\,.
\end{equation}
via the Harthree-Fock approximation \cite{ whitfield2011simulation, moll2016optimizing}.
The coefficients $h_{pq}$ and $h_{pqrs}$,
specific to each molecular system, 
respectively parametrize the one-body and 
two-body electron-electron interactions and can be efficiently computed classically.
The operators $c_p$ and $c_p^\dagger$ 
represent the fermionic annihilation 
and creation operators acting on an electron 
in the atomic orbital $p$.
The physical Hamiltonian (\ref{eq: ce{H2} Hamiltonian})
can be mapped onto $N$ qubits
via the Jordan-Wigner transformation \cite{shankar2017quantum},
given by the following equation:
\begin{equation}
    H = \Sigma_{i_1, ..., i_L} \left[a_{i_1, ..., i_L}\bigotimes_{j=1}^L\sigma^{i_j}\right],\,\,\, i_j \in \{0,1,2,3\} \, .
\end{equation}
Given the objective to create the  
$\ket{\Psi(0,1)}$ state,  
we take the auxiliary Hamiltonian to be 
the diagonal part of $H$
\begin{equation}
    H_0 = \mathrm{diag}\left(H\right).
\end{equation}
Since $H_0$ is diagonal,
the eigenstates are the elements of the computational basis,
making them easy to prepare on
a quantum computer.
In addition, each eigenstate can be directly associated with its corresponding eigenvalue.
The state $\ket{\Psi_0(0,1)}$ is prepared by
selecting a superposition comprising 
any ground state $\ket{\omega_0}$ and first excited state $\ket{\omega_1}$ of $H_0$. 
On the other hand,
the choice of the observable $O$ for small 
molecules is fully discussed in Appendix~\ref{appendix:observable}. 
These considerations 
result in an observable of the form
\begin{equation}
    O = \bigotimes_{j=1}^L \sigma^{i_j}_j \, , \quad i_j \in \{ 0, 1 \} \, ,
\end{equation}
which is a tensor product of identities and 
$\sigma^1$ matrices such that $\abs{\bra{\omega_1} O \ket{\omega_0}} = 1$.

We conducted numerical experiments to compute the ground-to-first-excited-state energy gaps for the diatomic molecules \ce{H2} and \ce{He2}. 
These systems were chosen as representative benchmarks to evaluate the performance and scalability of our approach.
When tackling molecular spectral gaps, 
instead of varying the transverse field as done for the Ising Hamiltonians, 
we test different internuclear distances between atoms, usually referred to as the \emph{bond length}.
For both experiments 40 Trotter steps
are used, using 5 adiabatic steps 
and 35 time evolution steps.
The \ce{H2} Hamiltonian requires 4 qubits,
while the \ce{He2} Hamiltonian requires 8 qubits.

\begin{figure*}[ht]
\centering
\begin{tikzpicture}
    \node[anchor=south west,inner sep=0] (H2plot) at (0,0)
        {\includegraphics[width=0.45\textwidth]{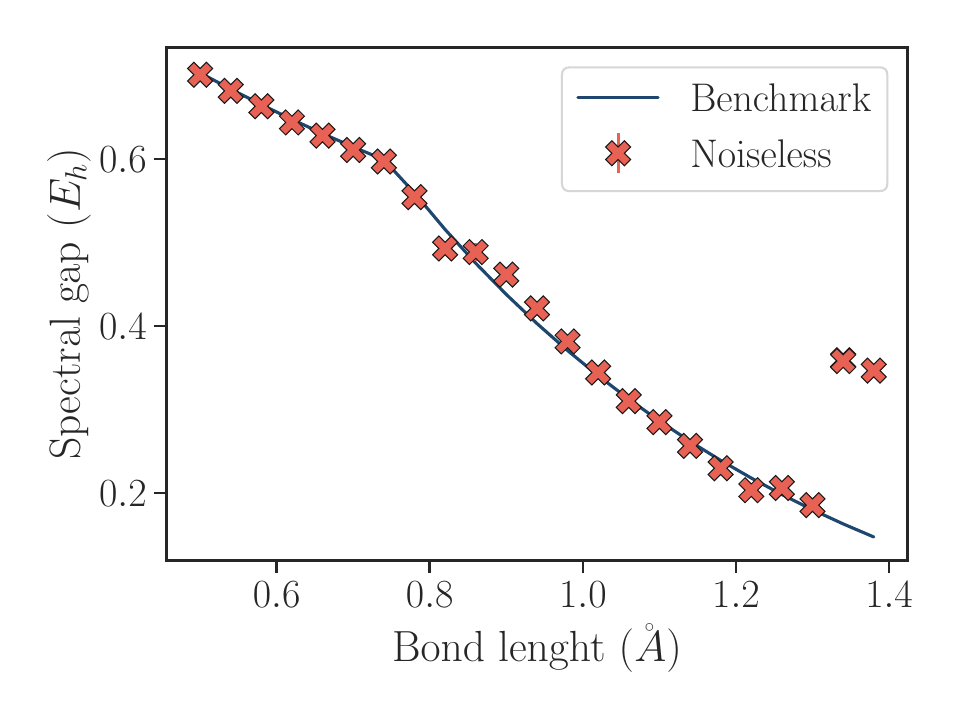}};
    \node[anchor=south west,inner sep=0] at ([xshift=1.8cm,yshift=1.8cm]H2plot.south west)
        {\includegraphics[width=2cm]{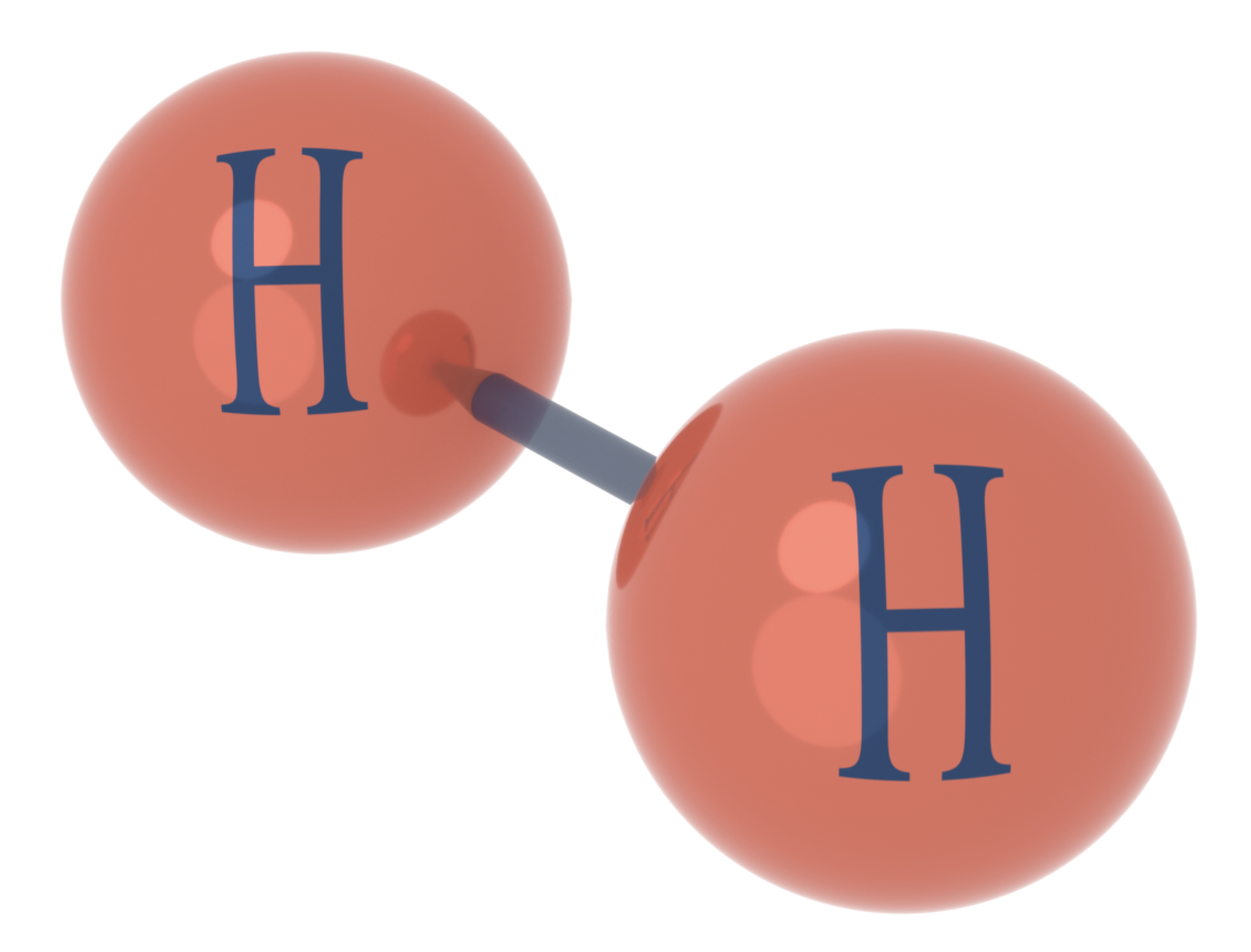}};
    \node[anchor=north west] at (H2plot.north west) {(a)};
\end{tikzpicture}
\hspace{0.05\textwidth} 
\begin{tikzpicture}
    \node[anchor=south west,inner sep=0] (He2plot) at (0,0)
        {\includegraphics[width=0.45\textwidth]{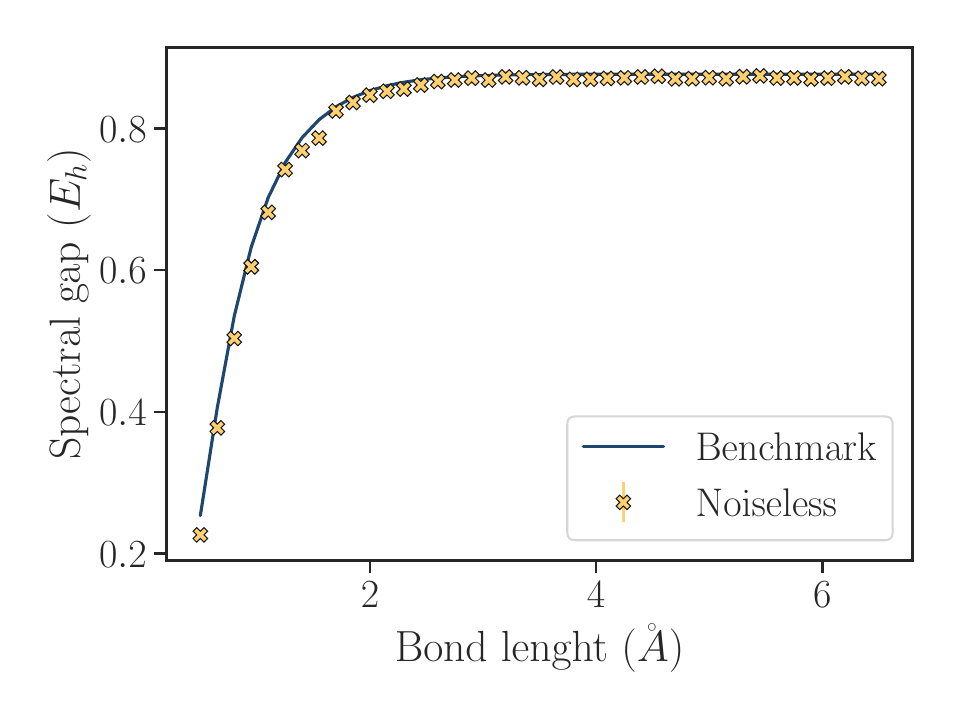}};
    \node[anchor=south west,inner sep=0] at ([xshift=5.5cm,yshift=2.8cm]He2plot.south west)
        {\includegraphics[width=2cm]{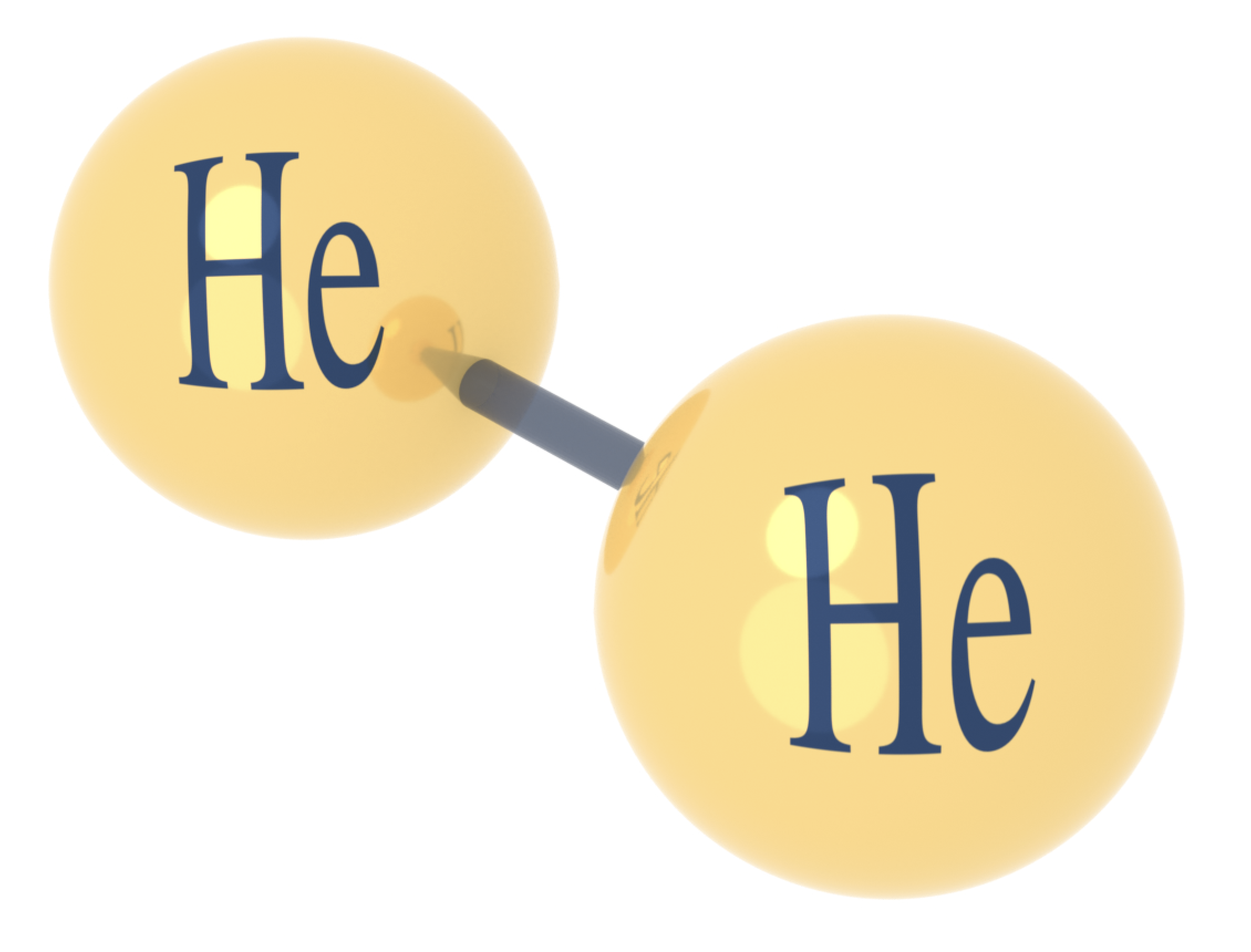}};
    \node[anchor=north west] at (He2plot.north west) {(b)};
\end{tikzpicture}

\caption{Spectral gap for (a) \ce{H2} and (b) \ce{He2} molecules as a function of the bond length. Noiseless simulations show excellent agreement with benchmark values. Error bars represent a single standard deviation computed from the fit. \label{fig: Molecules}}
\end{figure*}

Figure~\ref{fig: Molecules} shows the computed energy gaps for (a) \ce{H2} and (b) \ce{He2} as functions of bond length, compared against benchmark values obtained from exact diagonalization. In both cases, the noiseless simulations closely follow the benchmark curves, demonstrating high accuracy of the method across a broad range of geometries. 
Deviations are observed at larger bond lengths for \ce{H2}, 
due to imperfect state preparation. 
For \ce{He2}, the agreement remains excellent throughout, reflecting the robustness of the method even in cases with relatively flat potential energy surfaces.

Despite the procedure returning accurate
estimates of the spectral gap for a wide range of bond lengths,
as seen in Figure~\ref{fig: Molecules}, 
we believe that symmetry considerations would improve the feasibility of the state preparation on real hardware, e.g. by making the circuit shallower.
As a matter of fact, noisy simulations of molecular Hamiltonians remain impractical on current devices such as IonQ Aria 1, given present gate fidelities. From a counting argument, one can see that the number of 2-qubits gates required for state preparation (180 for \ce{H2} and 6640 for \ce{He2}) would result in prohibitively low overall fidelities
at the time of writing. 
\\

\section{Discussion}
\label{sec:discussion}
Bridging the gap between theoretical quantum algorithms and practical, near-term implementation requires tools that are both accurate and resource-efficient. Here, we presented a novel 
NISQ-friendly procedure to perform energy 
gap estimations for a wide range of Hamiltonians.
Our method improves upon previous works~\cite{GNATENKO2022127843} estimating the spectral gap via time evolution, by exploiting the Adiabatic Preparation method.
This enables to estimate any desired energy gap by 
preparing a tailored superposition state.
Our method is particularly well suited for estimating energy gaps in Hamiltonians with many non-commuting terms, as it avoids the significant overhead of conventional approaches, which require evaluating each Hamiltonian term for both eigenstates and subsequently computing their difference.
We demonstrated the capabilities 
of our spectral gap estimation procedure 
on the Ising model, both 1D and 2D, as 
well as small molecular systems, \ce{H2}
and \ce{He2}. 

We validated our method using circuit depths compatible with current NISQ devices, achieving relative errors on the order of $10^{-2}$ in noiseless simulations. 
The procedure remained accurate even with significantly reduced shot counts, highlighting its robustness.
Furthermore, we were able to 
accurately model the noise of the IonQ Aria 1 device to conduct noisy simulations.
This allowed the execution of experiments on the
real IonQ Aria 1 device and successfully estimate
the energy gap of 4-sites and 20-sites Ising chains
with a relative error of the order of $10^{-2}$. 

We note that the initial superposition state $\ket{\Psi(i,j)}$ cannot be straightforwardly prepared via the Variational Quantum Eigensolver (VQE). 
Furthermore, using AP instead of VQE, there is no risk of encountering problems such as barren plateaus, a fundamental limitation of variational algorithms which limits their applicability. 
However, it is important to note that AP does come with some caveats, such as requiring that the Hamiltonian has no energy level crossings, an issue often encountered in many-body physics or quantum chemistry. 
Consequently, 
any approach that relaxes these requirements 
can directly enhance the efficiency 
or broaden the applicability of our method, 
as demonstrated in our simulation of the Ising model using~\cite{cugini2025bsap}.
Given the rapid progress in quantum hardware, our procedure holds strong potential for applications to larger molecular systems and condensed matter models. 
This work paves the way for practical quantum computations on real quantum devices, with promising implications for both academic research and industrial applications in condensed matter physics and quantum chemistry.

\section{Data Availability Statement}
Code to reproduce the results and to create all figures presented in this manuscript is available at
Github repository~\cite{githubrepo}.

\section*{Acknowledgements}
We would like to thank 
Xanadu Quantum Technologies Inc for
organizing the Pennylane 2024 Coding
Challenges and Quantum Hackathon. 
The presented work was inspired by 
the Hackathon prompt ``Bridging the Gap''. 
Furthermore, we would like to thank 
Amazon Web Services for granting us 
Braket credits to run the experiments 
on real quantum hardware. 
F.G. would also like to acknowledge the high-performance computing cluster EOS at the Department of Mathematics, University of Pavia.
F.S., F.G. and A.R.M. were supported by the `National Quantum Science Technology Institute' (NQSTI, PE4) within the PNRR  project PE0000023.


\newpage
\appendix
\onecolumngrid

\section{Superposition states for Ising Hamiltonians}\label{Ising SGS}

We provide details on preparing the superposition state in \eqref{eq:SGS} for the Ising model with periodic boundary conditions (PBC). Our goal is to identify an initial state for the adiabatic preparation (AP) that evolves into the desired superposition state $\ket{\Psi}$ defined in Equation~\eqref{eq:SGS} as the final outcome.  
For our choice of $H_0$ (Eq.~\eqref{eq : initial Hamiltonina Ising}),
the ground state is degenerate,
and therefore the adiabatic theorem does not hold. 
To circumvent this issue we employ the Branched-Subspace Adiabatic Preparation (B-SAP) algorithm for state preparation~\cite{cugini2025bsap}, which extends the applicability of the adiabatic theorem to scenarios in which the initial Hamiltonian $H_0$ possesses a degenerate spectrum.
A direct implementation of B-SAP would require the application of a parametrized circuit prior to the evolution under $U_\tau$. 
However, for the specific case of the Ising model, 
the training of these parameters can be avoided by exploiting symmetry considerations.
To achieve this, we introduce the operator:  
\begin{equation}  
\mathcal{R} = \bigotimes_{j=0}^{L-1} R^{j}_z(\pi)\,,  
\end{equation}  
where $R_z^j (\theta)$ represents a rotation around the $z$-axis by angle $\theta$ on the $j$-th qubit. This operator commutes with the Hamiltonian for any value of $J_1$, ensuring that its expectation value remains conserved throughout the AP.  
The ground states of the auxiliary Hamiltonian $H_0$, $\ket{+}^{\otimes L}$ and $\ket{-}^{\otimes L}$, do not diagonalize $\mathcal{R}$, as:  
\begin{equation}  
\mathcal{R}|\pm\rangle^{\otimes L} = (-i)^L|\mp\rangle^{\otimes L}\,.  
\end{equation}  
However, by defining the new basis:  
\begin{equation}  
|\Phi^\pm\rangle = \frac{1}{\sqrt{2}}\left(|+\rangle^{\otimes L}\pm|-\rangle^{\otimes L} \right)\,,  
\end{equation}  
we obtain:  
\begin{equation}  
\mathcal{R}|\Phi^\pm\rangle = \pm(-i)^{L}|\Phi^\pm\rangle\,.  
\end{equation}  
Thus, the initial ground eigenspace is spanned by $\ket{\Phi^+}$ and $\ket{\Phi^-}$, which are eigenstates of $\mathcal{R}$ with eigenvalues $(-i)^L$ and $-(-i)^L$, respectively.

We now consider the state $\ket{\Psi(0,1)}$ defined in Equation~\eqref{eq:SGS}, which should be the equal superposition of the ground and first excited states of $H$, originating from the splitting of the initially degenerate ground eigenspace of $H_0$.  
In the limit $h_3/J_1 \to \infty$, the final ground state is:  
\begin{equation}  
\ket{\Omega_0}\equiv \ket{0}^{\otimes L}\,,  
\end{equation}  
with eigenvalue $(-i)^L$ with respect to $\mathcal{R}$. 
Since $\mathcal{R}$ is a conserved quantity, 
it follows that $\ket{\Omega_0}$ must evolve from $\ket{\Phi^+}$:  
\begin{equation}\label{eq: phi+ lim}  
\lim_{\tau \to \infty}U_\tau\ket{\Phi^+} =\ket{\Omega_0}\,. 
\end{equation}  
This also holds for any finite $h_3/J_1$, where a shorter adiabatic process is sufficient. 
By complementarity, we also obtain:  
\begin{equation}\label{eq: phi- lim}  
\lim_{\tau \to \infty}U_\tau\ket{\Phi^-} =\ket{\Omega_1}\,. 
\end{equation}  
Finally, by linearity, combining Eqs.~\eqref{eq: phi+ lim} and \eqref{eq: phi- lim}, we arrive at:
\begin{equation}    
    \lim_{\tau \to \infty}U_\tau \ket{+}^{\otimes L} =  \frac{1}{\sqrt{2}}\lim_{\tau \to \infty}U_\tau \left(\ket{\Phi^+}+\ket{\Phi^-}\right)   
    =\ket{\Psi(0,1)}\,.  
\end{equation}

\section{Observable choice}
\label{appendix:observable}
For our procedure to work,
the objective is to have an 
observable $O$ that maximizes  
$\mathcal{A}$, as described in Equation~\eqref{eq: fluctuations}. 
This facilitates a more straightforward 
estimation of the period of the sinusodial function, which in turn 
simplifies the fitting process 
for the energy gap. 

\subsubsection*{Ising Model}
\begin{figure}
    \centering
    \includegraphics[width=0.49\textwidth]{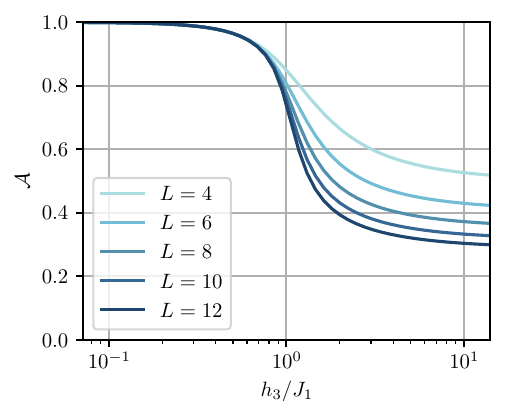}
    \caption{Numerical estimations of the expectation value $\mathcal{A} = \abs{\bra{\Phi^+}\sigma^1\otimes \left(\mathds{1}^{\otimes L-1}\right)\ket{\Phi^-}}$ as a function of the Ising's coupling constants ratio $J_1/h_3$.
    For all chain lengths considered, $L = \{4,6,8,10,12\}$, $\mathcal{A}$ remains larger than $L^{-1/2}$ for any value of $h_3/J_1$, ensuring that the amplitude is strictly positive across the entire range of coupling ratios.}   
    \label{fig:isingobv}
\end{figure}

For the Ising model our goal
is to find an operator $O$ for which $\mathcal{A}$ is substantially different from zero,
independently from the chain/lattice size.
At this scope we study the two regimes 
in which the eigenstates are analytically known.
In particular, for $h_3/J_1 \to 0$, one has
\begin{equation}
    \mathcal{A} = \abs{\bra{\Phi^+} O \ket{\Phi^-}}\,.
\end{equation}
Notably, any operator obtained with a tensor product 
of an odd number of $\sigma^1$ Pauli operations
gives $\mathcal{A} = 1$.
In the opposite regime, 
the one with $h_3/J_1 \to \infty$,
one has 
\begin{equation}
    \mathcal{A} = \abs{\bra{\Omega_0} O \ket{\Omega_1}} \\
    =  \frac{1}{\sqrt{L}}\abs{\sum_{j} \bra{0}^{\otimes L} O \ket{0}^{j} \otimes\ket{1} \otimes\ket{0}^{\otimes L-1-j} }\,.
\end{equation}
We therefore choose $O = \sigma^1_j\otimes \mathds{1}^{\otimes L-1}$,
where $j$ can be any site because of translational invariance,
so that $\mathcal{A} = 1/\sqrt{L} > 0$.
These limiting behaviors are illustrated in Figure~\ref{fig:isingobv}, where we plot $\mathcal{A}$ as a function of $h_3/J_1$ for different chain lengths.\\
Moreover,
we also test the values for $\mathcal{A}$ 
for all the possible choices of the observable $O$ with a Pauli-string shape 
on Ising chains with PBC 
and a length up to 10 sites.
Based on these considerations, 
we use the $\sigma^1_j$ operator
to estimate the energy gap of
all our experiments involving the
Ising model. 

\subsubsection*{Molecules}
For small molecules, the choice of the
observable $O$ is based on
the following considerations. 
Since both ground and excited state of 
the auxiliary Hamiltonian are in 
the computational basis,
it is always possible to find a 
tensor product $P$ of identities and 
$\sigma^1$ Pauli matrices such that
\begin{equation}
    \bra{\omega_1} P \ket{\omega_0} = \bra{\omega_1} \bigotimes_{j=1}^L \sigma^{i_j}_j \ket{\omega_0} = 1\,, \quad i_j \in \{ 0, 1 \}\,.
\end{equation} 
Assuming that the off-diagonal terms of $H$ are small with respect to the diagonal terms, then choosing $P$ as the operator for our procedure will satisfy $\mathcal{A}\neq0$.
In particular, for both \ce{H2} and \ce{He2} molecules, we choose the operator $P$ such that:
\begin{equation}
    \mathcal{A} = \abs{\bra{\Omega_1} P \ket{\Omega_0}} = \abs{\bra{\omega_1} U_\tau^\dagger P U_\tau\ket{\omega_0}} \, .
\end{equation}

\section{Quantum measurement error}\label{sec: Intrinsic error}
Let 
\begin{equation}
\langle O \rangle = \langle \psi |O|   \psi \rangle  
\end{equation}
be the expectation value 
of an operator $O$
on a quantum state $\ket{\psi}$,
and let 
\begin{equation}
\langle\left( O - \langle O \rangle \right)^2 \rangle
\end{equation}
be its variance.
Recalling that $O$ should be unitary and Hermitian 
in order to be measured on a quantum computer 
\begin{equation}
O^2 = O^\dagger O = \mathds{1}\,,
\end{equation}
so that its eigenvalues can only be $\{+1, -1\}$.
In the non trivial case in which 
the eigenvalues are different,
one can then decompose the state 
onto two eigenstates of $O$ as
\begin{equation}
|\psi\rangle  = \alpha|\psi_{+1}\rangle + \beta|\psi_{-1}\rangle \, ,
\end{equation}
with
\begin{equation}
\begin{cases}
O|\psi_{\pm1}\rangle = \pm |\psi_{\pm1}\rangle\\
|\alpha|^2+ |\beta|^2 = 1\,.
\end{cases}
\end{equation}
Then
\begin{equation}
\begin{cases}
\langle O \rangle = |\alpha|^2-|\beta|^2 \\
    \langle\left( O -  \langle O \rangle \right)^2 \rangle = 4|\alpha|^2|\beta|^2 \, .\\
\end{cases} 
\end{equation}

\section{Imperfect superposition state preparation}
\label{appendix:imperfect sgs}

In this Appendix, 
we provide insights into the robustness of the proposed method with respect to imperfect superposition state preparation. 
We henceforth only consider the effects of 
the finite time for both Adiabatic Preparation (AP)
and of the Trotterization process,
without taking into account 
those coming from quantum noise.
In particular, we will now show analytically that if the error in the preparation is of order $\epsilon$ then the estimation of the expectation value will experience an error of the same order. 

Let  
\begin{equation}\label{eq:approximates_SGS}
    \ket{\nu_{ij}(\epsilon)} = \ket{\Psi_{ij}} + \mathcal{O}(\epsilon) 
    = \frac{1}{\sqrt{2}}\left( \ket{\Omega_i} + \ket{\Omega_j} \right) 
    + \mathcal{O}(\epsilon)
\end{equation}
be an imperfectly prepared superposition state,  
where $\epsilon$ denotes the error magnitude.  
This state can be expressed on the basis of the Hamiltonian eigenstates as
\begin{equation}\label{eq:series_expansion_error}
    \ket{\nu_{ij}(\epsilon)} = \sum_n c_n(\epsilon) \ket{\Omega_n}\,.
\end{equation}
We expand the coefficients in powers of $\epsilon$:
\begin{equation}\label{eq: c_n series}
    c_n = \sum_{m=0}^\infty c_n^{(m)} \epsilon^m\,.
\end{equation}
The coefficients $c_n$ can be uniquely determined by imposing the same conditions typically used in perturbation theory (see, e.g., \cite{landau2013quantum}).  
First, we require that for $\epsilon = 0$, the ideal state is recovered, i.e. $\ket{\nu_{ij}(0)} = \ket{\Psi_{ij}}$.  
This condition holds if and only if
\begin{equation}
    c_i^{(0)} = \frac{1}{\sqrt{2}}\,, 
    \quad c_j^{(0)} = \frac{1}{\sqrt{2}}\,, 
    \quad c_{n \neq i,j}^{(0)} = 0\,,
\end{equation}
or, more compactly,
\begin{equation}
    c_n^{(0)} = \frac{1}{\sqrt{2}} \left( \delta_{n i} + \delta_{n j} \right)\,.
\end{equation}
As a second condition, we require that the first-order truncated state be normalized up to second-order corrections, namely
\begin{equation}
     1  - \braket{\nu_{ij}(\epsilon)}{\nu_{ij}(\epsilon)}= \mathcal{O}(\epsilon^2)\,.
\end{equation}
Substituting Eq.~\eqref{eq:series_expansion_error} 
and Eq.~\eqref{eq: c_n series}
one obtains
\begin{equation}
    1 -\Bigg[
        \frac{1}{\sqrt{2}}(\bra{\Omega_i} + \bra{\Omega_j})
        + \epsilon \sum_n c_n^{(1)\,*} \bra{\Omega_n}
        + \mathcal{O}(\epsilon^2)
    \Bigg] 
     \times
    \Bigg[
        \frac{1}{\sqrt{2}}(\ket{\Omega_i} + \ket{\Omega_j})
        + \epsilon \sum_n c_n^{(1)} \ket{\Omega_n}
        + \mathcal{O}(\epsilon^2)
    \Bigg] = \mathcal{O}(\epsilon^2)\,,
\end{equation}
which yields $c_i^{(1)} = c_j^{(1)} = 0$.  
Therefore,
\begin{equation}
    \ket{\nu_{ij}(\epsilon)} 
    = \frac{1}{\sqrt{2}} \left( \ket{\Omega_i} + \ket{\Omega_j} \right)
    + \epsilon \sum_{n \neq i,j} c_n^{(1)} \ket{\Omega_n}
    + \mathcal{O}(\epsilon^2)\,.
\end{equation}

The expectation value of a time-dependent operator $O(t)$ on $\ket{\nu_{ij}(\epsilon)}$ is then
\begin{equation}
\begin{split}
    \bra{\nu_{ij}(\epsilon)} O(t) \ket{\nu_{ij}(\epsilon)}
    = \bra{\Psi_{ij}} O(t) \ket{\Psi_{ij}} 
    &+ \epsilon \sqrt{2} \sum_{n \neq i,j}
    \Big[
        \abs{c_n^{(1)} \bra{\Omega_i} O \ket{\Omega_n}}
        \cos(\Delta_{i,n} t + \theta_{i,n}) \\
    &
        + \abs{c_n^{(1)} \bra{\Omega_j} O \ket{\Omega_n}}
        \cos(\Delta_{j,n} t + \theta_{j,n})
    \Big]
    + \mathcal{O}(\epsilon^2)\,.
\end{split}
\end{equation}

\begin{figure}
    \centering
    \includegraphics[width=0.49\textwidth]{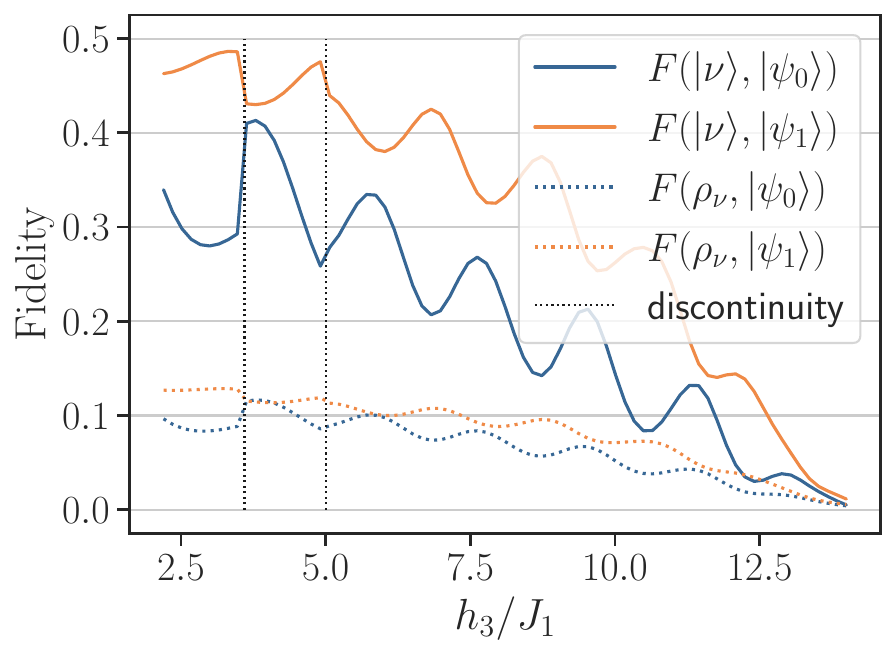}
    \caption{Fidelity of the approximate superposition state $\ket{\Psi(0,1)}$, in the noiseless and noisy case ($\ket{\nu}$, ${\rho_\nu}$), with the ground and first excited state ($\ket{\psi_0}$, $\ket{\psi_1}$) for an Ising chain of $L=8$ qubits. The fidelity is found to be sub-optimal, decaying and oscillating for increasing $h_3$. The discontinuities are given by the change in $\delta \tau$ (at $h_3=\{3.6,5\}$) imposed in the adiabatic process.}
    \label{fig:fidelity_oscillations}
\end{figure}

Finally, we relate the error magnitude $\epsilon$ to the infidelity, defined as
\begin{equation}
    \mathcal{I} = 1 - \abs{\braket{\nu_{ij}(\epsilon)}{\Psi_{ij}}}^2
    = \mathcal{O}(\epsilon^2)\,.
\end{equation}
From this relation, we conclude that if the target state $\ket{\Psi_{ij}}$ is prepared with a given infidelity $\mathcal{I}$, 
the estimated expectation value of a generic time-dependent operator $O(t)$ is affected by an error proportional to $\sqrt{\mathcal{I}}$.

In particular, we numerically verify the fidelity of the prepared superposition state with both the ground and first excited states of the one-dimensional Ising chain as a function of the transverse field, as shown in Figure~\ref{fig:fidelity_oscillations}. 
The theoretically expected optimal fidelities of the superposition state with the ground and first excited states are 
\begin{equation}
    F(\ket{\Psi},\ket{\psi_0}) = F(\ket{\Psi},\ket{\psi_1}) = 0.5\,.
\end{equation}
As illustrated in Figure~\ref{fig:fidelity_oscillations}, the fidelity does not reach the optimal theoretical value and decreases with increasing $h_3$, both in the noiseless and in the noisy simulations. 
Consistently with the theoretical analysis presented in this Appendix, even though the prepared state $\ket{\Psi}$ is imperfect, the proposed method remains robust: it yields accurate estimates of expectation values with errors on the order of $10^{-2}$ and provides a reliable approximation of the spectral gap of the system. 
Finally, we note the presence of oscillations in the fidelity, which may originate from intrinsic features of the adiabatic preparation process, even though we were not able to identify them theoretically.

\section{More details on IonQ Aria}
\label{appendix:ionq}
\subsubsection{IonQ Native Gates}
\label{sec:native gates}
\begin{figure}[]
    \centering
    \includegraphics[trim={1.cm 1cm 1.9cm 1cm},clip,width=.48\textwidth]{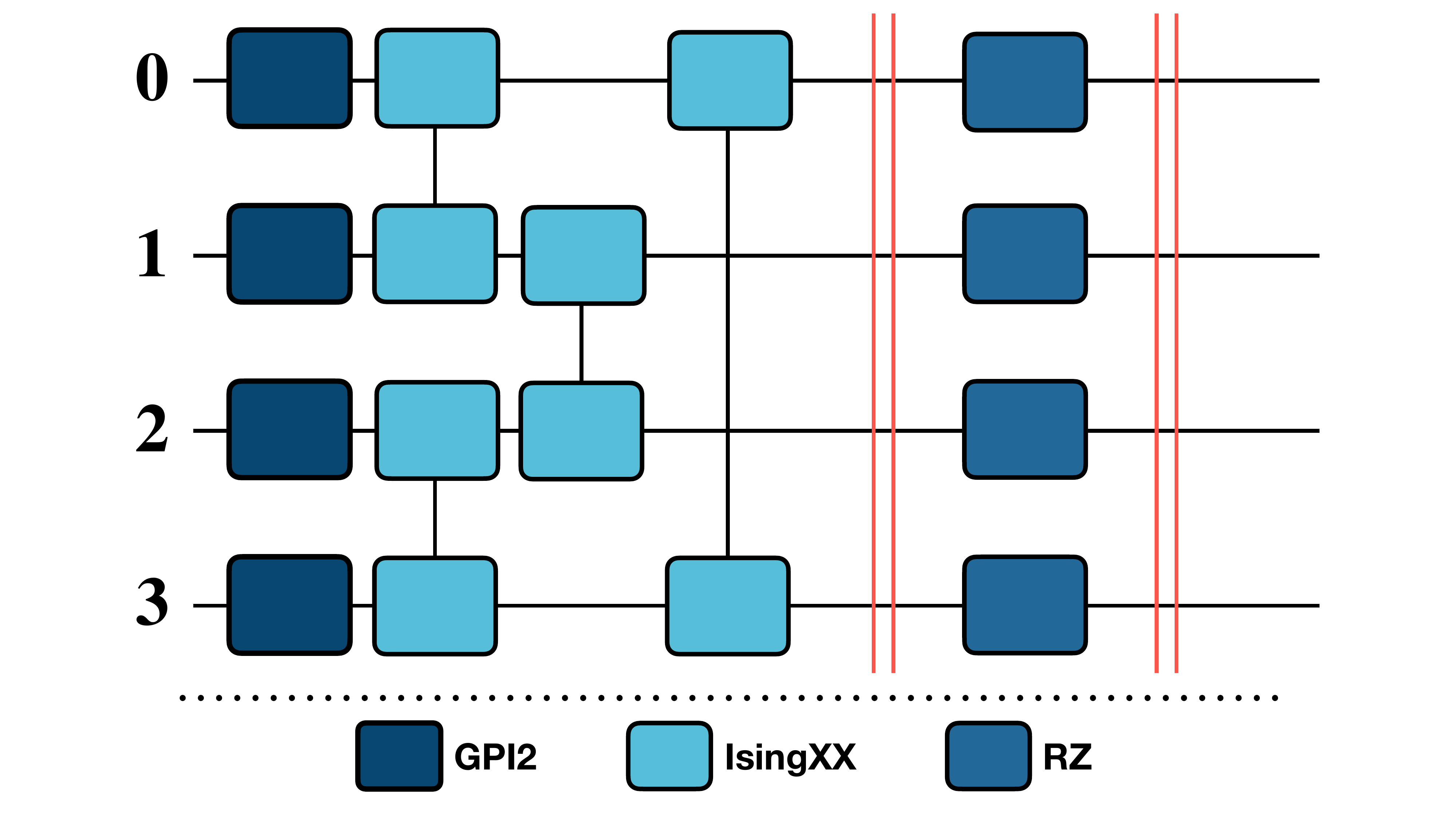}
    \caption{Quantum circuit for a single Trotter step showing the evolution of the superposition state $\ket{\Psi}$. The circuit involves the application of GPI2($\phi$), RZ($\theta$), and IsingXX($\theta$) gates to prepare $\ket{\Psi}$ for spectral gap estimation.}
    \label{fig:trotterstep}
\end{figure}

Since single qubit-gates are typically implemented on ion-trapped hardware with high average gate fidelity, potential issues in result's fidelity (when running a circuit on NISQ hardware) primarily stem from 2-qubit gates. For this reason, we simulate the Ising model with interactions along the $x$-axis and transverse field along the $z$-axis. This allows us to use only the native entangling gates available for IonQ Aria device, i.e. Mølmer-Sørensen gate (MS)~\cite{sorensen1999quantum}, and to reduce the number of entangling gates only to the essential ones. More precisely, in a single Trotter step, one only needs two layers of entangling gates (see Figure~\ref{fig:trotterstep}), implying 80 2-qubit gates depth when fixing the number of steps to 40.

The native gates are 
\begin{equation}
    \text{GPI2} (\phi) = \frac{1}{\sqrt{2}} \begin{pmatrix}
        1 & -ie^{-i\phi} \\
        -ie^{i\phi} & 1
    \end{pmatrix},
    \label{eq:gpi2}
\end{equation}
\begin{equation}
VirtualZ (\theta) = \begin{pmatrix}
        e^{-i\frac{\theta}{2}} & 0 \\
        0 & e^{i\frac{\theta}{2}}
    \end{pmatrix},
    \label{eq:virtualZ}
\end{equation}
and 

\begin{equation}
       \text{MS}(\phi_0, \phi_1, \theta) = \begin{pmatrix}
        \cos{\frac{\theta}{2}} & 0 & 0 & -ie^{-i(\phi_0+\phi_1)}\sin{\frac{\theta}{2}}\\
        0 & \cos{\frac{\theta}{2}} & -ie^{-i(\phi_0-\phi_1)}\sin{\frac{\theta}{2}} & 0\\
        0 & -ie^{i(\phi_0-\phi_1)}\sin{\frac{\theta}{2}} & \cos{\frac{\theta}{2}} & 0\\
        -ie^{i(\phi_0+\phi_1)}\sin{\frac{\theta}{2}} & 0 & 0 & \cos{\frac{\theta}{2}}
    \end{pmatrix}.
\label{eq:MS}
\end{equation}
The Mølmer-Sørensen gate can be further simplified to obtain the Ising interaction $\sigma_i^1\sigma_j^1$ by setting $\phi_0$ and $\phi_1$ equal to 0.

\subsubsection{IonQ Aria noise model}
\label{sec:ionq noise}
To assess the resilience of our approach to noise, we incorporated a realistic noise model taking into account the technical features of IonQ's Aria device available online~\cite{ionq_aria}. 
The noise sources considered include:
\begin{itemize}
    \item [-] \textit{Thermal Relaxation}: This source of noise accounts for the effects of the thermal relaxation times $T_1$ and $T_2$. Notably, IonQ Aria has $T_1 = 100s$ and $T_2=1s$, and while we took them into account in our simulations, their effects were found to be negligible and did not significantly impact the results.
    \item [-] \textit{Gate Fidelity}: We simulated errors arising from finite gate fidelity, modeled as depolarizing noise. The relationship between gate fidelity and the depolarizing parameter is derived from what reported in the Appendix of~\cite{blank2020quantum}. The two-qubit gate time is $600\mu s$, while single qubit gates are executed in $135\mu s$.
    \item [-] \textit{Readout Noise:} A readout noise of 0.39\% was introduced to emulate imperfections in the measurement process.
\end{itemize}
Upon detailed examination, we have determined that the gate fidelity plays a significant role in the estimation of the spectral gap. This noise source has been implemented by considering a single qubit depolarizing channel, which acts on a density matrix $\varrho$ 
\begin{equation}
    \mathcal{D}_p (\varrho) = (1-p) \varrho + p\frac{\mathds{1}}{2}\varrho \ .
    \label{eq:depolarizing}
\end{equation}
To take into account gate imperfections and finite qubit relaxation times we define the probability $p$ as
\begin{equation}
    p = 1 + 3 \frac{2\epsilon - 1}{d} \ .
\end{equation}
Here, $d =\exp{-T_g/T_1} + 2\exp{- T_g/T_2}$ and $\epsilon = 1- F$, where $F$ is the average gate fidelity. The term $T_g$ represents the gate time, which is the duration required to perform a single-qubit or two-qubit gate operation. We define two depolarizing channels one per each type of gate (single/two-qubit gate). 

\begin{figure}
    \centering
    \includegraphics[width=.5\textwidth]{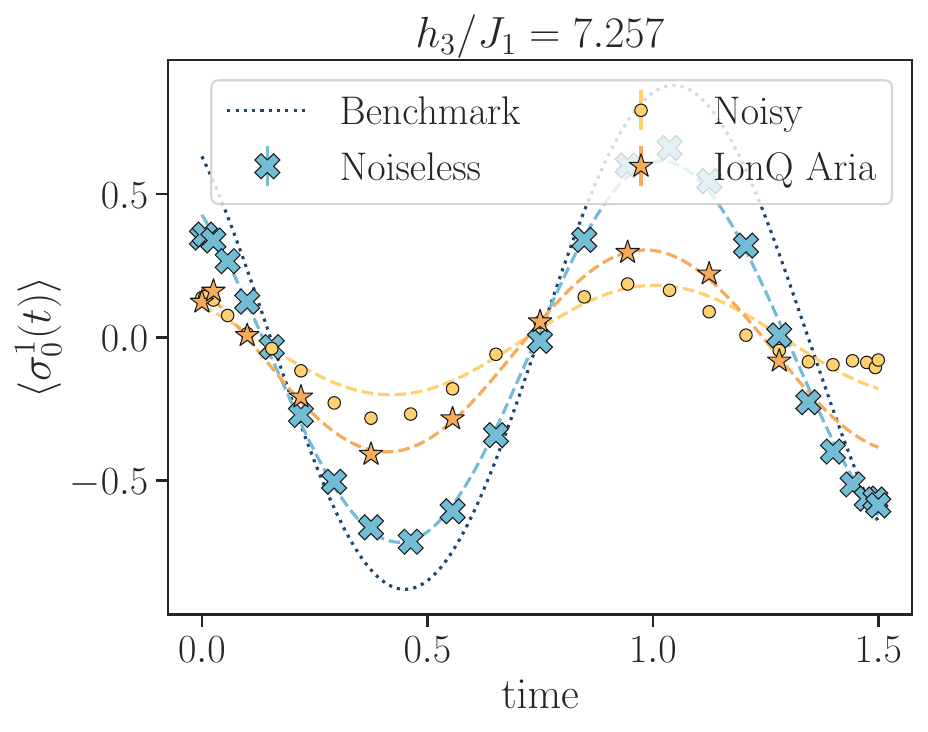}
    \caption{Observable oscillations in time for Ising 1D chain with $L=4$ at $h_3/J_1 = 7.257$, measured (points) and fitted (dashed lines) for noiseless, noisy and real hardware (with error mitigation). For reference, it is also reported the expected oscillatory from diagonalization (dotted line). Error bars represent a single standard deviation computed from the fit.}
    \label{fig:waves}
\end{figure}

Our approach demonstrates robustness against gate fidelity errors and readout noise, suggesting its viability for simulation on real quantum hardware.

The results of the noise resilience analysis are presented in Figure \ref{fig:Ising}. The robustness of our approach can be appreciated also in the time oscillations of the observable reported in Figure~\ref{fig:waves}. There, the damping of oscillations due to noise (manifested as a reduction in $\mathcal{A}$) is highlighted, compared to the expected behaviour $\mathcal{A} \cos\left(\Delta E+\theta\right)$. The amplitude $\mathcal{A}$ is taken from our preliminary study of the observable, and the phase $\theta$ is set as in the noiseless simulations. Interestingly the noise mitigation allows for having wider oscillations on real hardware compared to noisy simulations.

\begin{figure}[t]
    \centering
    \includegraphics[width=.45\textwidth]{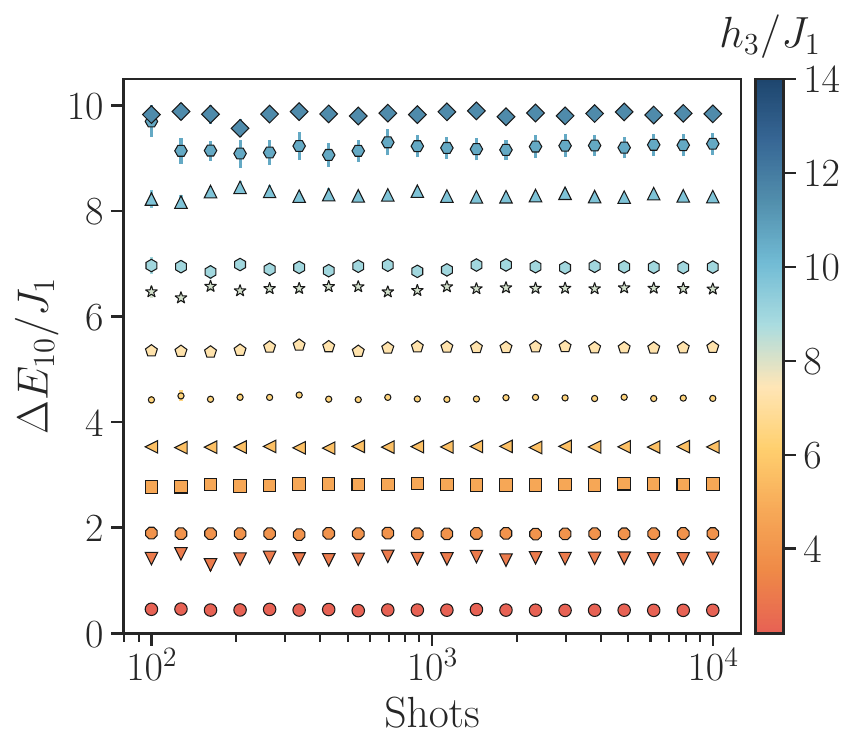}
    \caption{Robustness analysis as a function of the number of shots for the 1D Ising chain with $L = 4$. Scatter points show the measured spectral gap in units of $J_1$, and error bars represent one standard deviation obtained from the fit.}
    \label{fig:shots noise}
\end{figure}

In particular, the IonQ Aria device reduces the impact of systematic errors by employing a specific error mitigation method called \textit{debiasing}~\cite{maksymov2023enhancing}. This technique maps a circuit into multiple variants, employing different qubit permutations or gate decompositions. The effectiveness of debiasing error mitigation is likely improved by our approach's robustness to a reduced number of shots. In fact, as reported in Figure~\ref{fig:shots noise}, we assess the robustness of our approach over a reduced number of shots by reducing this value from $10^4$ up to 100 shots for chains of 4 qubits. As witnessed by Figure~\ref{fig:shots noise}, the spectral gap estimate stays approximately constant over different number of shots.

\begin{figure*}[]
    \centering
    \includegraphics[trim={.cm 0.cm 0cm 0cm},clip,width=.45\textwidth]{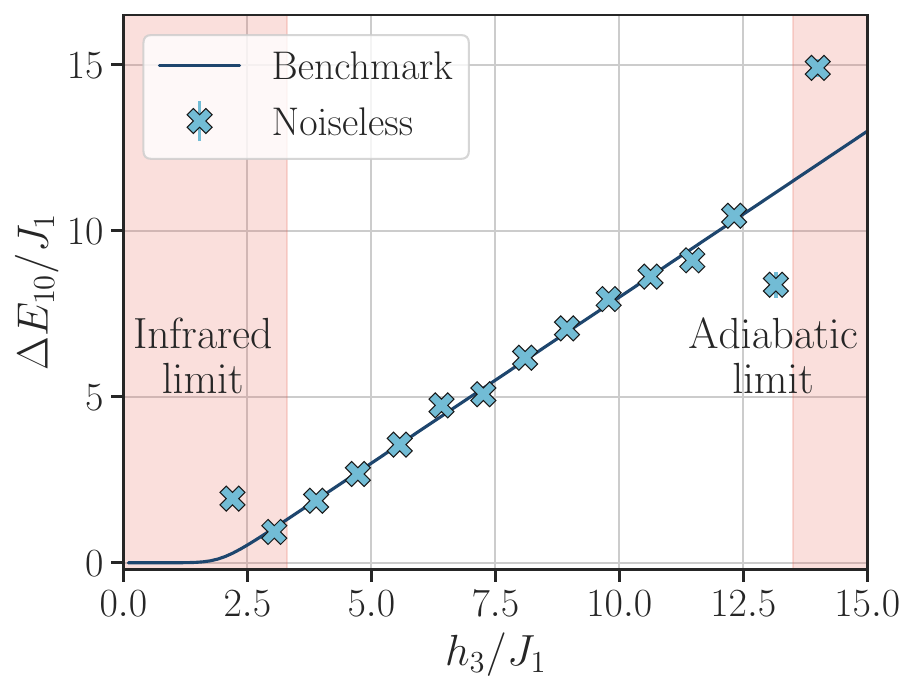}
    \shiftleft{8.4cm}{\raisebox{5.3cm}[0cm][0cm]{(a)}}
    \label{fig:spectralgaprunning4}
    \includegraphics[trim={.0cm 0.cm 0.cm 0cm},clip,width=0.425\textwidth]{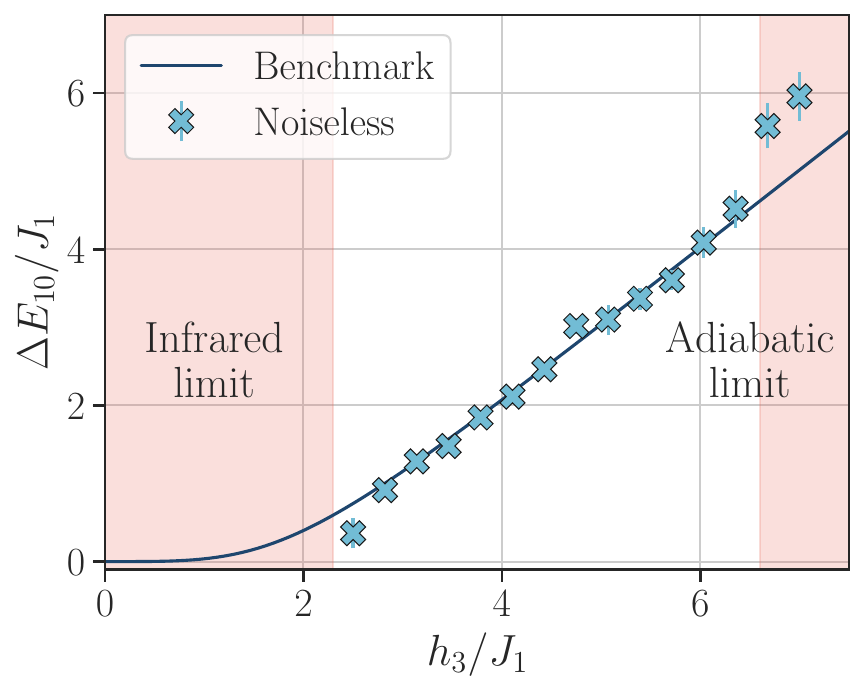}
    \shiftleft{8.2cm}{\raisebox{5.3cm}[0cm][0cm]{(b)}}
    \caption{Spectral gap as a function of $h_3/J_1$ for a) a $L=10$ Ising 1D and b) a $L=4$ Ising 2D lattice with periodic boundary conditions (PBC). Our approach demonstrates optimal agreement with the benchmark values, in the noiseless case. Error bars represent a single standard deviation computed from the fit.}
    \label{fig:Ising 2D}
\end{figure*}

\section{Additional Ising experiments}
\label{subsec:Ising2d}

Here, we extend our study to two additional Ising configurations with periodic boundary conditions—a 10-spin chain and a 4-site lattice. Their spectral-gap dependence on $h_3/J_1$ is presented in Figure~\ref{fig:Ising 2D}. All previous conditions for the Ising chain
experiment remain in this study. In particular, the 40 Trotter steps are divided in 15 adiabatic steps and 25 time evolution steps. 

The numerical results for the 10-site Ising chain show excellent agreement between the benchmark values and the noiseless estimate of the spectral gap. We do not report experiments on real hardware because the limited availability of the IonQ Aria device prevented additional tests. Simulations with a realistic noise model were also not performed in this case, as we consider the proof-of-principle demonstrations with 4 and 20 qubits presented in the main text to be sufficient.
In the 2D simulations, the number of entangling gates required per Trotter step is doubled with respect to the 1D model. Due to the prohibitive depth implied by the extra interactions, simulations with real hardware noise model lead to inconclusive results. It would be interesting to see if IonQ Aria noise mitigation is able to lead to measurable results.

\bibliography{bibliography}

\end{document}